# Real-time observation of epitaxial graphene domain reorientation

Paul C. Rogge[1,2], Konrad Thürmer[3], Michael E. Foster[3], Kevin F. McCarty[3], Oscar D. Dubon[1,2,†], and Norman C. Bartelt[3,*]

[1]Department of Materials Science and Engineering, University of California, Berkeley, Berkeley, CA 94720
[2]Materials Sciences Division, Lawrence Berkeley National Laboratory, Berkeley, CA 94720
[3]Sandia National Laboratory, Livermore, CA 94550
*bartelt@sandia.gov
†oddubon@berkeley.edu

Abstract: Graphene films grown by vapor deposition tend to be polycrystalline due to the nucleation and growth of islands with different in-plane orientations. Here, using low-energy electron microscopy, we find that micron-sized graphene islands on Ir(111) rotate to a preferred orientation during thermal annealing. We observe three alignment mechanisms: the simultaneous growth of aligned domains and dissolution of rotated domains, i.e., "ripening"; domain-boundary motion within islands; and continuous lattice-rotation of entire domains. By measuring the relative growth velocity of domains during ripening, we estimate that the driving force for alignment is on the order of 0.1 meV per C atom and increases with rotation angle. A simple model of the orientation-dependent energy associated with the moiré corrugation of the graphene sheet due to local variations in the graphene-substrate interaction reproduces the results. This work suggests new strategies for improving the van der Waals epitaxy of 2D materials.

     Attempts to grow large, structurally pristine two-dimensional (2D) materials such as graphene and transition-metal dichalcogenides are often frustrated by the fact that different rotational variants nucleate and grow, leading to a polycrystalline film[1-5]. This is not surprising given the weak film-substrate interactions inherent in the van der Waals epitaxy of these materials, which disables typical epitaxial control such as lattice matching. It is surprising, however, that rotational disorder in graphene films greatly varies with the substrate. For example, a single orientation prevails on Ru(0001)[6] and, under optimized growth conditions, on Ge(110)[7]. Graphene on Cu spans a small range of orientations[2], whereas graphene on Pd(111)[8] and Ir(111)[4,9-10] can adopt many orientations. What then controls island orientation? The relevant factors are not understood. Consequently, current efforts to improve film quality focus on reducing nucleation density[11-13] or using substrate-specific recipes such as selective etching steps[14]. A more general approach requires understanding what dictates island orientation during van der Waals epitaxy.

     Here, we use low-energy electron microscopy (LEEM) to observe in *real-time* the evolution of graphene domains during thermal annealing. We show that large, micron-scale graphene domains on Ir(111) can reorient themselves to a preferred orientation after nucleation by three mechanisms. We estimate that domains aligned with the Ir(111) lattice are energetically favorable by as little as ~0.1 meV per C atom. A simple model of the orientation-dependent energy associated with the moiré corrugation of the graphene sheet due to local variations in the graphene-substrate interaction reproduces the results. These results provide valuable guidance for synthesis of 2D materials on current and future substrates: annealing immediately after

nucleation will improve quality and this effect will be greater for substrates that induce larger corrugations in the 2D sheets.

**Results**

**Three mechanisms of graphene domain reorientation.** The first mechanism identified, shown in Fig. 1a,b, is that some orientations grew while others shrank simultaneously. Figure 1c compares the two frames and highlights the changes after 155 minutes. R0 and R11 domains grew while the R12 domain simultaneously dissolved. The second process appears upon further annealing. After 200 minutes the reflectivity of the R12 domain circled in Fig. 1b changes in time. This transformation occurs over ~80 minutes as shown by the recorded reflectivity in Fig. 1e. Selected-area low-energy electron diffraction (LEED) (Fig. 1f) reveals that this entire region has undergone a continuous change in orientation from R12 to R4. Interestingly, the initial R12 domain dissolves as the newly created R4 domain grows. The third process is shown in Fig. 1g. As the first two processes occur, the boundary between the R11 and R12 domain, denoted by the red arrows, move into the R12 domain (see also Supplementary Movie 1). We performed 20 such annealing experiments at constant temperature in which these domain reorientation mechanisms were observed. More highly rotated domains always dissolved while less-rotated domains grew. Within those 20 experiments, the continuous lattice rotation was observed five times and domain boundary motion was observed only once. Typically, an hour-long annealing step eliminated half of the rotational variants.

We now turn to the cause of the simultaneous growth and dissolution. In general, growth or dissolution is dictated by the surface adatom concentration adjacent to the island, $c$, relative to the equilibrium adatom concentration for the graphene domain, $c_{eq}$. Growth occurs when $c > c_{eq}$ and dissolution occurs when $c < c_{eq}$. For an ideal adatom lattice gas, the chemical potential difference associated with growth or dissolution is $\Delta\mu \approx kT \ln c/c_{eq}$, where $k$ is the Boltzmann constant and $T$ is the absolute temperature[15]. LEEM reflectivity measurements show that the adatom concentration around each island is the same throughout the regions of observation (tens of microns), which is consistent with attachment-limited growth of graphene on Ir[10,16]. Therefore, the simultaneous growth of a domain of one orientation and dissolution of another with a different orientation indicates that the equilibrium adatom concentration of a given domain depends on its orientation relative to the substrate and, correspondingly, that the chemical potential is orientation dependent as illustrated in Fig. 2a.

The relative chemical potential of different rotations can be immediately determined by combining the experimental results, as shown in Fig. 2b. All observations indicate that domains aligned with the Ir substrate, R0, have the lowest chemical potential and that the chemical potential increases as the angle of rotation increases from R0. This picture is consistent with the observed spontaneous change in domain orientation shown in Fig. 1d. A domain can lower its chemical potential by reducing its rotation angle. With its new orientation, the R4 domain has a lower $c_{eq}$ compared to the R12 domain, which results in the growth of the R4 domain after the transformation while the remaining R12 domain continues to dissolve.

**Temperature-dependent ripening**. The simultaneous growth and dissolution, i.e., "ripening", is sensitive to temperature. As one sees in Fig. 3, graphene islands were imaged as the temperature was varied in steps of ~5 K. At lower temperatures, all domains grew, while at high temperatures domains shrank. By adjusting the temperature, we could control which domains grew and which domains dissolved. This temperature dependence is due to segregation of dissolved C in the Ir bulk to the surface when the amount of C dissolved in the bulk becomes greater than the solubility limit. At lower temperatures, bulk segregation causes the carbon adatom chemical potential to be above the equilibrium value of all islands. As the temperature increases, the segregation flux to the surface decreases and the adatom concentration decreases below the equilibrium value.

What is the magnitude of the driving force for a rotated domain RX to realign itself to R0? To estimate it, we use the temperature-controlled ripening experiments described above to estimate the differences in equilibrium adatom concentrations between different rotations. From these adatom concentrations, we then extract the difference in chemical potentials of the islands. In previous work, the velocity of R0 domains was precisely determined as a function of the adatom concentration and temperature[10] using the equation[17]

$$v_{R0} = B(T) \left[ \left( \frac{c}{c_{eq,R0}} \right)^n - 1 \right] \qquad (1)$$

where $c_{eq,R0}$ is the equilibrium adatom concentration for the R0 island, $n$ is the cluster size for attachment to a kink and is equal to four, and $B(T) = \exp[31 - (33800 \pm 1300)/T]$ nm/s is the kinetic coefficient[10]. Converting the $A(t)$ plot of Fig. 3c into a growth velocity allows us to determine the adatom concentration when the R14 and R8 domains are stationary and in equilibrium with the adatoms. If a rotated domain, RX, is stationary while a standalone R0 island continues to grow, the difference in chemical potential can be estimated from the adatom concentrations by

$$\mu_{RX} - \mu_{R0} \approx kT \ln \frac{c_{eq,RX}}{c_{eq,R0}} \approx kT \ln \left[ \left( \frac{v_{R0}}{B} + 1 \right)^{1/n} \right]. \qquad (2)$$

When the R14 domain was stationary, the R0 velocity was $540 \pm 36$ nm hr$^{-1}$ (1000 °C), where the error is from the standard deviation of the d$A$/d$t$ line fit. When the R8 domain was stationary from 2800-3200 seconds (1030 °C), the R0 velocity decreased to $144 \pm 36$ nm hr$^{-1}$. Using equation (2) we find that the R14 domain has a chemical potential of $0.07 \pm 0.01$ meV per C atom above R0 while R8 is $0.01 \pm 0.01$ meV per C atom above R0.

Five temperature-controlled experiments were conducted in the same manner. The estimated differences in chemical potential relative to R0 are shown in Fig. 3d. The results were reproducible across the five experiments and did not depend on the shape or size of the islands (see Supplementary Note 1). Thus, we attribute the differences in chemical potential to differences in binding strength of the various rotations. The chemical potential increases with rotation from R0, as expected. The trend is not linear, with small differences in chemical potential for small rotations and larger increases in chemical potential as the rotation angle increases. Due to the uncertainty in $B(T)$, the absolute scale of the chemical potential may be a factor of two larger or a quarter smaller.

These measurements allow us to estimate the difference in chemical potential that cause the observed grain boundary motion in Fig. 1g. It is extremely small, 0.007 meV per C atom. An alternative, albeit less direct, argument based on the boundary curvature induced by edge pinning gives a similar estimate (see Supplementary Note 2). We note that we only observed the boundary move in the case of one degree difference in rotation: evidently domain boundary mobility must decrease with relative misorientation and limit this mechanism of coarsening. On the other hand, the continuous rotation was seen five times and was active when grain boundary motion was not.

**Moiré corrugation model.** These small energy differences seem beyond the capabilities of ab-initio techniques such as density functional theory (DFT) to compute directly. To understand the origin of the orientation-dependent chemical potential differences and the extent to which they can be generalized, we make a model of the energy differences due to the corrugation of the graphene sheet, which is known to significantly change with orientation. In incommensurate, undistorted moirés, all equivalent positions of C with respect to the substrate are occupied with equal probability, independent of the layer's rotation[18], suggesting an equal energy for all rotational variants. Changing lattice misfit with the substrate does not alter this conclusion; so it is doubtful that lattice misfit determines the preferred orientation. However, it is experimentally observed that graphene sheets are not undistorted. Instead, experimental results[4,19-21] and DFT[22] have shown that graphene sheets on metals are corrugated. Scanning tunneling microscopy (STM) images of R14, R19[9], and R30[4] graphene on Ir(111) reveal that, in addition to the large-scale modulation with moiré periodicity, the graphene topography also exhibits a pronounced fine-scale corrugation (in the case of R30 with a periodicity of ~5 Å), which is absent in the R0 phase. We have used DFT to compute the structure as a function of orientation (see Supplementary Fig. 1 for the optimized structures). As seen in Fig. 4a we reproduce these experimental trends.

The reason for this corrugation is that the preferred distance of the graphene sheet from the substrate varies locally, depending on the lateral position of the C atoms with respect to the substrate (e.g., whether the C atom is at an atop position, a threefold hollow site, etc.). To maximize its binding to the substrate, the graphene sheet flexes slightly in order to follow a contour of minimum C-Ir potential. Due to the bending rigidity of the graphene, the carbon atoms are further away from their optimal positions as the wavelength of the corrugation shortens. So the short-wavelength corrugations observed experimentally for rotated graphene would lead to an energy cost.

To estimate the magnitude of this effect, we construct a simple model for the observed corrugations. For simplicity, we assume that the graphene is incommensurate with respect to the Ir substrate, i.e., that the possible energy reduction due to locking into a (higher order) commensurate lattice is not sufficient to overcome the energy cost of straining the graphene from its preferred lattice constant. We assume that the preferred distance $z_0$ of carbon atoms from the Ir(111) substrate is represented by a two-dimensional sinusoid which has opposite extrema at the atop sites and hollow sites. We take the force on each carbon atom to be proportional to how far the atom is displaced from the preferred distance and assume that the spring constant $k$ is uniform. In the relaxed configuration, this force is balanced by the forces caused by bending the graphene sheet. In the continuum limit, the energy cost $E_b$ of periodic corrugations is[23,24]

$$E_b = \frac{1}{2} \lambda \int K^2 \mathrm{d}S \qquad (3)$$

where $\lambda$ is the mean bending rigidity, ~1.4 eV[24], and $K$ is the mean curvature. Our model approximates the local curvature $K_i$ at a given atom $i$ as being proportional to the height difference $\Delta z_i$ between atom $i$ and the average of its surrounding three nearest neighbors. This gives a force on each atom equal to $3\sqrt{3}\lambda \Delta z_i/d^2$, where $d$ is the C-C distance in graphene (for a derivation, see Supplementary Note 3). The equilibrium sheet corrugation is given when this bending force balances the displacement force.

This model has only two parameters, the amplitude of the sinusoidal preferred C atom distance and the spring constant $k$. To apply this model to graphene on Ir(111), we fit these two parameters to our DFT calculations for R0 shown in Fig. 4a. DFT is reliable for R0 corrugations because it reproduces those from experiments[22]. From the DFT calculations we obtain the equilibrium corrugation and the forces on each C atom for a sheet constrained to be flat. We then adjust the amplitude of the sinusoidal preferred separation distance and the spring constant $k$ to exactly reproduce the amplitude of the corrugation of the relaxed sheet and the root mean square of the forces on the flat sheet. Given these parameters, we predicted the corrugation for domains rotated by 7.5, 15, and 30 degrees. Height profiles extracted from our moiré corrugation model along the close-packed zigzag direction are shown in Fig. 4b.

The agreement between this prediction and DFT calculations is striking. It reproduces the position of all C atoms in the profiles in Fig. 4b to within 0.05 angstroms, with most atoms within 0.01 angstroms. The good agreement makes it plausible that it correctly captures the energy due to C atom displacement and bending of the graphene sheet with changing orientation. And indeed, as shown in Fig. 4c, despite its extreme simplicity, the model correctly reproduces the observed stability of R0 and it easily accounts for the magnitude of the measured energy differences. That the model tends to overestimate the energy differences is not surprising given that the model does not allow lateral relaxation of atomic positions and the factor of two in the uncertainty of the scale of the experimental data (Fig. 3d). (Such small energy differences are generally difficult to compute directly with DFT because of the large moiré unit cells; however, we have used DFT to estimate the binding energy difference between R30 and R0 and find it to be less than ~1 meV per C atom, consistent with our model and experiment. See Methods and Supplementary Table 1.) Thus the driving forces for the post-nucleation reorientations that we observe can be explained by the orientation dependence of the moiré corrugation. This driving force applies to all three observed reorientation mechanisms.

**Discussion**

Our results have significant implications on the nature of rotational disorder of graphene islands on various substrates. STM shows that graphene on Au(111)[3] and Cu(111)[25] exhibits the same trend of increasing nearest-neighbor corrugation with rotation angle as seen on Ir(111). One would expect that the R0 orientation to be energetically preferred. As expected, like on Ir(111)[16], the majority orientation on Cu(111)[2] and Au(111)[26] is indeed R0. This is especially noteworthy for Au(111), where R0 has a lattice mismatch of 17%[26] and clearly supports the conclusion that the moiré corrugation dictates the resulting island orientation. This raises the possibility that in the systems where good alignment is observed in as-grown films, e.g., Cu(111)[2], Ru(0001)[6], Au(111)[26], and Ge(110)[7], such alignment occurs shortly after nucleation by the continuous lattice rotation mechanism when the islands are still small. Furthermore, these results imply that the rotational disorder of other 2D materials (hexagonal-BN, transition-metal dichalcogenides) can be controlled only when grown on substrates where corrugations exist, e.g., on metals[27-30].

In summary, we have identified the energetically preferred graphene orientation on Ir(111). Domains can evolve to this orientation during thermal annealing by three distinct mechanisms. The driving force for the alignment can be as small as 0.1 meV per C atom, much less than the total binding strength of graphene to metal surfaces. We propose that the origin of the preferential alignment is caused by the varying degree to which carbon atoms can attain the preferred distance from the substrate: the graphene bending rigidity prevents the sheet from following the short wavelength corrugations inherent in highly rotated domains. Since graphene corrugations of similar size are ubiquitous, the driving force for the coarsening on Ir reported here is expected to be similar in other systems. Despite this small driving force for rotational alignment, our results indicate that post-nucleation annealing of 2D materials can still improve rotational order. This strategy means that high quality 2D materials can be grown on more substrates than is now recognized. Our understanding of the driving force for reorientation can guide the search for new substrates.

## Methods
### Experimental
Graphene islands were nucleated by heating the Ir substrate between 800-900 °C and introducing ethylene gas into the LEEM chamber. After nucleation, the ethylene flow was turned off. The temperature was measured by a W-Rh thermocouple welded directly to the substrate.

### Determining domain orientation
The domain orientation relative to the Ir(111) lattice was identified by selected-area low-energy electron diffraction. The moiré LEED spots are sensitive to small changes in domain orientation and were used to determine the orientation to within ±0.5°. The distance between each pair of moiré superstructure LEED spots was measured and normalized by the distance between the corresponding pair of first-order graphene LEED spots. This ratio was computed for the three pairs of spots per LEED pattern and then averaged. Over 47 LEED patterns were measured and orientations were assigned based on this ratio. This ratio is easily determined using a simple vector model and there was excellent agreement between the experimental and expected ratios.

### Velocity measurement
The average R0 growth velocity was computed by measuring its island area, $A$, and perimeter, $p$, with time: $v_{\text{ave}} = (1/p_{\text{ave}})dA/dt$ where $dA/dt$ is the slope of a line fit to the area vs. time data for the time interval in question, and $p_{\text{ave}}$ was the average perimeter over the time interval.

### Density Functional Theory
All DFT calculations were performed with the optB86b-vdW functional with spin-polarization and using the projector-augmented-wave (PAW) GW-pseudopotentials supplied by VASP. A cutoff energy of 400 eV for the plane-wave basis and the Brillouin zone was sampled using 1x1x1 and 2x2x1 k-point grids. The optB86b-vdW exchange-correlation functional was employed in order to account for van der Waals interactions that are required for a qualitative description of the intermolecular interactions between the graphene sheet and the Ir(111) surface. The functional is of the form: $E_{\text{xc}} = E_{\text{x}}^{\text{optB86b}} + E_{\text{c}}^{\text{LDA}} + E_{\text{c}}^{\text{vdW-DF}}$, where $E_{\text{x}}^{\text{optB86b}}$ is the exchange energy, $E_{\text{c}}^{\text{LDA}}$ is the local density approximation (LDA) to the correlation energy, and $E_{\text{c}}^{\text{vdW-DF}}$ is the non-local correlation energy term.

The Ir(111) surface was constructed from the optimized face-centered cubic structure, $a = b = c = 3.843$ Å; the lattice of the surface was fixed during all simulations. The Ir(111) slab consisted of three atomic layers, which helped minimize computational cost and allowed for the full graphene moiré to be modeled. Five different graphene orientations were considered: R0

(542 atoms: 10x10 Ir supercell; 300 Ir and 242 C atoms), R7.5 (776 atoms: 12x12 Ir supercell; 432 Ir and 344 C atoms), R15 (437 atoms: 9x9 Ir supercell; 243 Ir and 194 C atoms), R22.5 (344 atoms: 8x8 Ir supercell; 192 Ir and 152 C atoms), and R30 (657 atoms: 11x11 Ir supercell; 363 Ir and 294 C atoms). Geometry optimizations were performed until all forces were less than 0.05 eV Å$^{-1}$; the total energy was converged to within 1x10$^{-6}$ eV for both k-point samplings. The binding energies per carbon atom of the systems were determined by:

$$E_{\text{BE}} = \frac{E_{\text{G@Ir111}} - E_{\text{G}} - E_{\text{Ir111}}}{N}, \quad (8)$$

where $E_{\text{G@Ir111}}$ is the total energy of the system (R0, R7.5, R15, R22.5, or R30 on the Ir(111) substrate), $E_{\text{G}}$ and $E_{\text{Ir111}}$ are the total energies of the optimized isolated graphene layer and the Ir(111) surface, and $N$ is the total number of carbon atoms. The graphene deformation energies (bending energies) were determined from the difference between the single-point energy of the optimized graphene sheet on the Ir(111) surface and the fully optimized graphene sheet within the same unit cell.

Two k-point grids were used to compute graphene deformation energies and binding energies per C atom. The larger 2x2x1 k-point grid was used to test whether the predicted energies from the 1x1x1 k-point grid were converged with respect to k-space. The results indicate that the computed graphene deformation energies are converged, however, the binding energies are not. This conclusion was reached by comparing the difference in energies with respective to k-space (ΔE/Δk). For the deformation energies, ΔE/Δk is much smaller than the magnitude of the values. On the other hand, the relative binding energies between the different orientations (R0, R7.5, R15, R22.5, and R30) are smaller than ΔE/Δk. Based on this analysis, it is concluded that the predicted ordering of the deformation energies is statistically significant, but the relative binding energies between orientations is not significant. As a result, the orientation with the greatest binding energy per C atom cannot reliably be determined from these results. Increasing k-space much further is computationally intractable; moreover, each system has different lattice dimensions that affect the relative k-point spacing. Computationally determining differences of one-tenth of a meV is difficult given the various numerical approximations used within DFT (i.e., k-points and energy cutoffs).

6788-6796 (2014).

**Acknowledgements**
This work was supported by the Director, Office of Science, Office of Basic Energy Sciences, Division of Materials Sciences and Engineering, of the U.S. Department of Energy Contract No. De-Ac04-94AL85000 (SNL) and by the NSF under Grant No. DMR-1105541 (ODD, PCR).

**Author contributions**
P.C.R., O.D.D., K.F.M., and N.C.B. conceived the experiments. P.C.R. performed the LEEM experiments and K.F.M. assisted. P.C.R., N.C.B., and K.T. conceived the moiré corrugation model and K.T. developed and implemented the model. M.E.F. performed DFT calculations. All authors participated in the data discussion and writing of the manuscript.

**Competing financial interests**
The authors declare no competing financial interests.

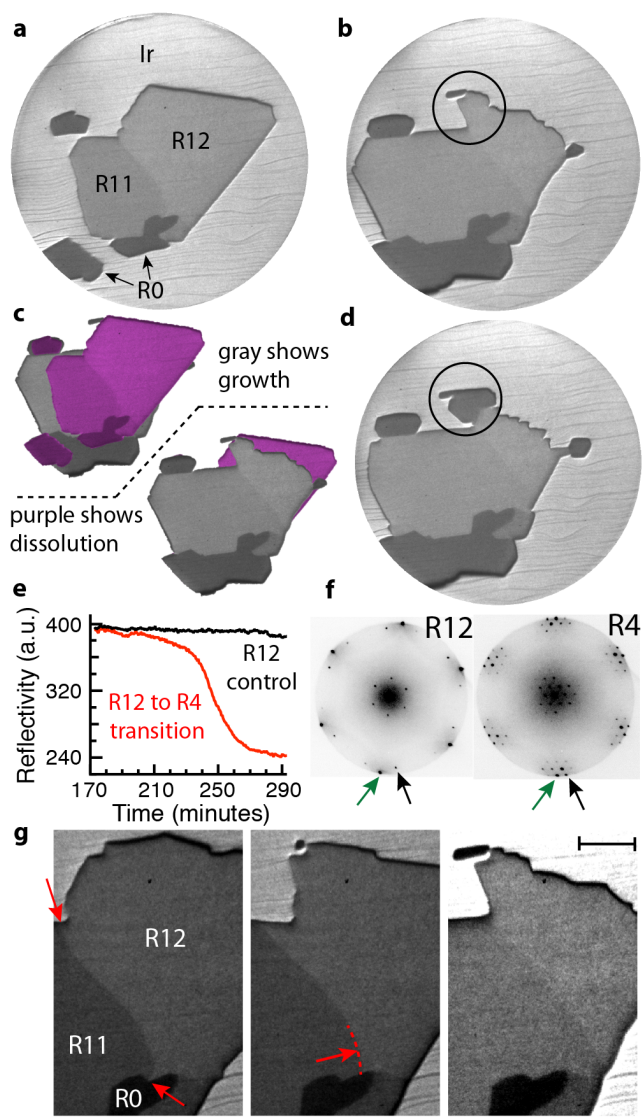

**Figure 1 Three domain realignment mechanisms of graphene on Ir(111)**. **a** LEEM image of a sub-monolayer graphene film on Ir(111) at 1050 °C (75 µm field of view (FOV)). The spatial extent of individual domains of a given orientation can be deduced from differences in the reflectivity within the graphene islands and its rotation angle is denoted by the number following the letter R, as determined by selected-area low-energy electron diffraction (LEED). **b** Evolution of graphene islands after annealing at 1050 °C for 155 minutes. **c** Image stacks of **a** and **b** after removing the background shows how the islands evolved. With the initial frame (false-colored purple) on top, exposed areas show growth, reversing the image order exposes areas that dissolved. **d** LEEM image after annealing for 257 minutes; circles in **b** and **d** mark a section of the R12 domain that continuously transformed to the R4 orientation. **e** Island reflectivity monitored with time for the R12 domain and the R12 to R4 transition region. **f** Selected-area LEED taken before and after the reflectivity change. Arrows: graphene (green) and Ir(111) (black) first-order LEED spots. **g** Domain boundary motion between R11 and R12. Middle frame highlights the radius of curvature of the boundary as it moved (red arrow). Time, from left to right: 55, 102, 165 minutes. Scale bar: 10 µm.

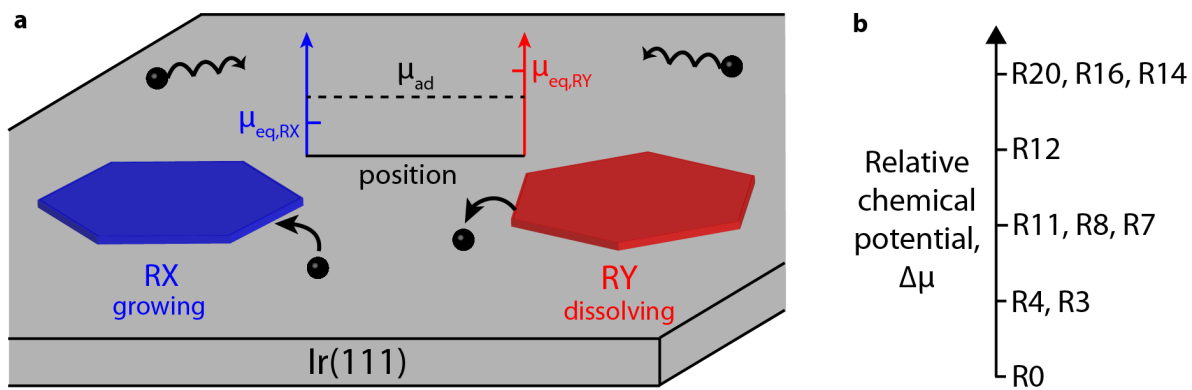

**Figure 2 Orientation-dependent chemical potential as determined from simultaneous growth and dissolution experiments. a** Interpretation of the simultaneous growth and dissolution of graphene domains, RX and RY, respectively, for a spatially uniform C adatom concentration. **b** Relative chemical potential of graphene rotational variants. Multiple rotations on one line indicate that these rotation couples have not been observed together in the same experiment and cannot be further distinguished.

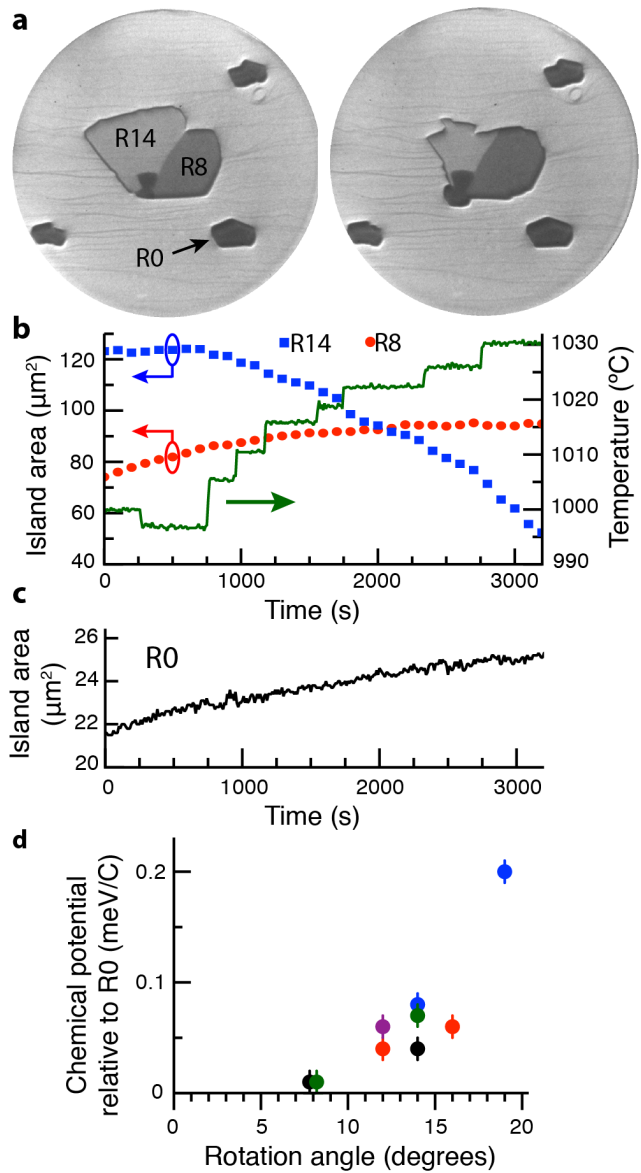

**Figure 3 Temperature-controlled ripening of graphene domains**. **a** LEEM images (50 μm FOV) of graphene islands at 500 seconds (left) and 3200 seconds (right) during temperature-controlled ripening. **b** R14 and R8 domain areas measured every 10$^{th}$ frame (every 100 seconds). At 997 °C (200-800 seconds), all domains grew; increasing the temperature to 1005 °C (800-1000 seconds) resulted in the R14 domain dissolving while the R8 domain continued to grow. Increasing the temperature further ended the simultaneous dissolution and growth between the R14 and R8 domains: at 1030 °C the R8 domain ceased growing (2800-3200 seconds). Substrate temperature (green line) was adjusted in steps of ~5 K. **c** Standalone R0 island area was measured every frame (every 10 seconds) and continued to grow the entire time. **d** Quantitative differences in chemical potential relative to R0. Individual experiments are color-coded; R8 data were offset in the x-axis for clarity. Error bars are from the uncertainty in the velocity measurement of R0 islands used to determine the difference in chemical potential.

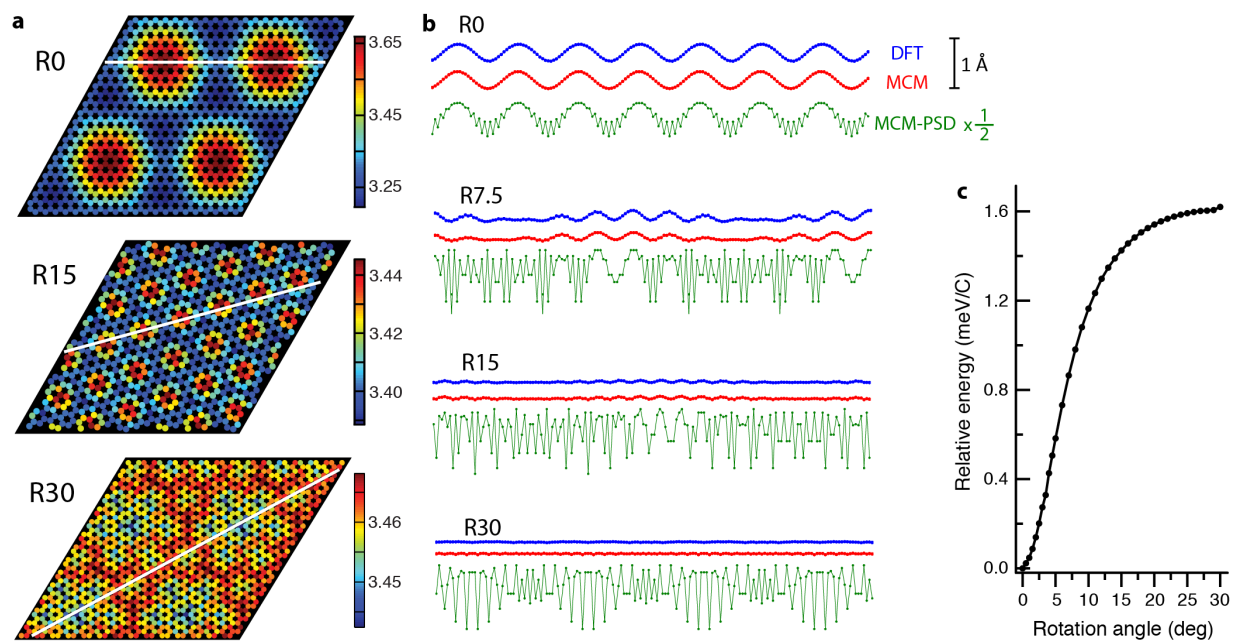

**Figure 4 Moiré corrugation model**. **a** Distance of C atoms from the Ir(111) substrate obtained from DFT calculations for graphene rotated by 0, 15, and 30 degrees relative to the Ir(111) lattice. Scale bar is in angstroms. **b** Moiré corrugation model (MCM) results. Green line: the preferred separation distance of C atoms from the substrate (MCM-PSD) taken along the close-packed zigzag direction (representative white lines in **a**), where filled circle represent C atoms. Due to the bending rigidity of the graphene sheet, the C atoms cannot follow the short-wavelength corrugations and the resulting sheet corrugation from the MCM is given by the red line. Corrugation profiles taken from the DFT results are shown in blue for comparison. Scale bar applies to all three profiles for all four rotations. Note: the MCM-PSD profile is multiplied by ½ for clarity. **c** Energy per C atom relative to the R0 orientation as a function of rotation angle as determined from the moiré corrugation model.

**Supplementary Note 1**

One might be concerned that edge energies would be important in determining the chemical potential differences, as they are in standard Ostwald ripening. However, we observed no influence of R0 island size on ripening velocity, which suggests that edge energies are not a significant factor. The R0 areas varied in size from 17 μm$^2$ to 90 μm$^2$ among the various experiments. For example, in one experiment, a R14 domain was stationary at 1010 °C, while the R0 velocity was 0.20 nm/s and had an area of 76 μm$^2$. In another experiment, a R14 domain was stationary at 1000 °C, while the R0 velocity was 0.15 nm/s and had an area of 22 μm$^2$. Their velocity ratio was ~1.3; if the velocity was influenced by edge energies, the velocity ratio should depend on the ratio of the island radii, ~1.9. Instead, the velocities can be accounted for by the ratio of the kinetic coefficients, $B$(1010 °C)/$B$(1000 °C) ≈ 1.2. This indicates that the R0 velocity was independent of island size for these island sizes.

**Supplementary Note 2**

We further support the reported differences in chemical potential by analyzing the curvature of the moving domain boundary shown in Fig. 1g in the main manuscript. Initially, the boundary has a structure in which the curvature changes sign, i.e., the curvature inverts, an artifact attributed to the initial growth of the island. When the boundary began to move, the curvature changed such that the boundary extended into the R12 domain only (Fig. 1g, middle frame). This curvature is attributed to pinning by the lower R0 domain. Indeed, as the bottom end of the boundary moved beyond the R0 domain, the boundary straightened (Fig. 1g, right frame). The radius of curvature of the boundary while it was moving, but partially pinned, was ~9 μm, as seen in the middle frame of Fig. 1g. We use this curvature to estimate the difference in chemical potential between the two domains.

Consider an idealized situation in which two phases with different chemical potentials meet at a boundary with both ends pinned. The boundary will move in response to the chemical potential difference, but because its ends are pinned, it will become curved and resultantly lengthen. Equilibrium is achieved when the free energy cost of further increasing the boundary length is equal to the free energy gain by transferring mass from the phase of higher to lower chemical potential, in which the difference in chemical potential is given by

$$\Delta \mu = \frac{\Omega \gamma}{R} \qquad (1)$$

where $\Omega$ is the atomic area, $\gamma$ is the domain boundary energy per unit length, and $R$ is the radius of curvature. Although the boundary is moving in the example shown in Fig. 1g, we use the equilibrium picture to estimate a lower bound of the difference in chemical potentials, i.e., if the boundary was pinned completely, the radius of curvature would decrease, and thus increase $\Delta\mu$. *Ab-initio* calculations previously estimated $\gamma$ as a function of misorientation between two graphene domains. For a misorientation of 1°, we extrapolate the results from Supplementary Reference 5, which gives $\gamma \approx 0.05$ eV/Å. Given the graphene atomic density of $3.8*10^{15}$ cm$^{-2}$ and the 9 μm radius of curvature between the R11 and R12 domain, the resulting difference in chemical potential is ~0.002 meV/C (Supplementary equation (1)). As a comparison, the velocity analysis reported in Fig. 3d gives $\Delta\mu$ = 0.007 meV between R11 and R12. As expected, this result is higher than that obtained from the domain boundary motion. More importantly, the domain boundary analysis supports the order of magnitude of the reported values in Fig. 3d.

**Supplementary Note 3**

Consider the curvature around atom $i$ and assume atom $i$ and its three nearest-neighbors (NN) lie on a paraboloid (~sphere with radius $r = 1/k$ where $k$ is the curvature) with a coordinate system centered at atom $i$ and the x-y plane tangential to the paraboloid. Consider everything to be rather flat and approximate lengths by their projections, $z = (1/2)(k_1 x^2 + k_2 y^2)$ and assume $k_1 = k_2 = k = 1/r$. Let the $x$-axis point to a neighboring atom's projection. This neighbor (and all other nearest neighbors) are higher than atom $i$ by

$$\Delta h_i = z_i = \frac{1}{2} k_i d^2 \tag{2}$$

with C-C distance $d$ = 1.42 Å. The curvature at atom $i$ is

$$k_i = 2z_i/d^2 \text{ or } k_i^2 = 4z_i^2/d^4. \tag{3}$$

From continuum theory, the energy due to bending is[6]

$$E_b = \frac{1}{2} \lambda \int K^2 dS \tag{4}$$

where $\lambda$ is the mean bending rigidity, ~1.4 eV[7], $K$ is the mean curvature, and noting that the Gaussian bending vanishes for a 2D, periodic sheet[7]. Converting the integral to a sum over $i$ gives

$$E_b = \frac{1}{2} \lambda \sum k_i^2 S_0 \tag{5}$$

where the footprint of a C atom $S_0 = 3d^2 \sqrt{3}/4$. Substituting for $k_i$ (Supplementary equation (3)) and $S_0$,

$$E_b = \frac{3\sqrt{3}\lambda}{2} \sum (z_i/d)^2, \tag{6}$$

or the bending energy per atom

$$\frac{E_b}{N} = \frac{3\sqrt{3}\lambda}{2N} \sum (z_i/d)^2. \tag{7}$$

Differentiating with respect to $z_i$ gives the expression for the force on each atom specified in the text

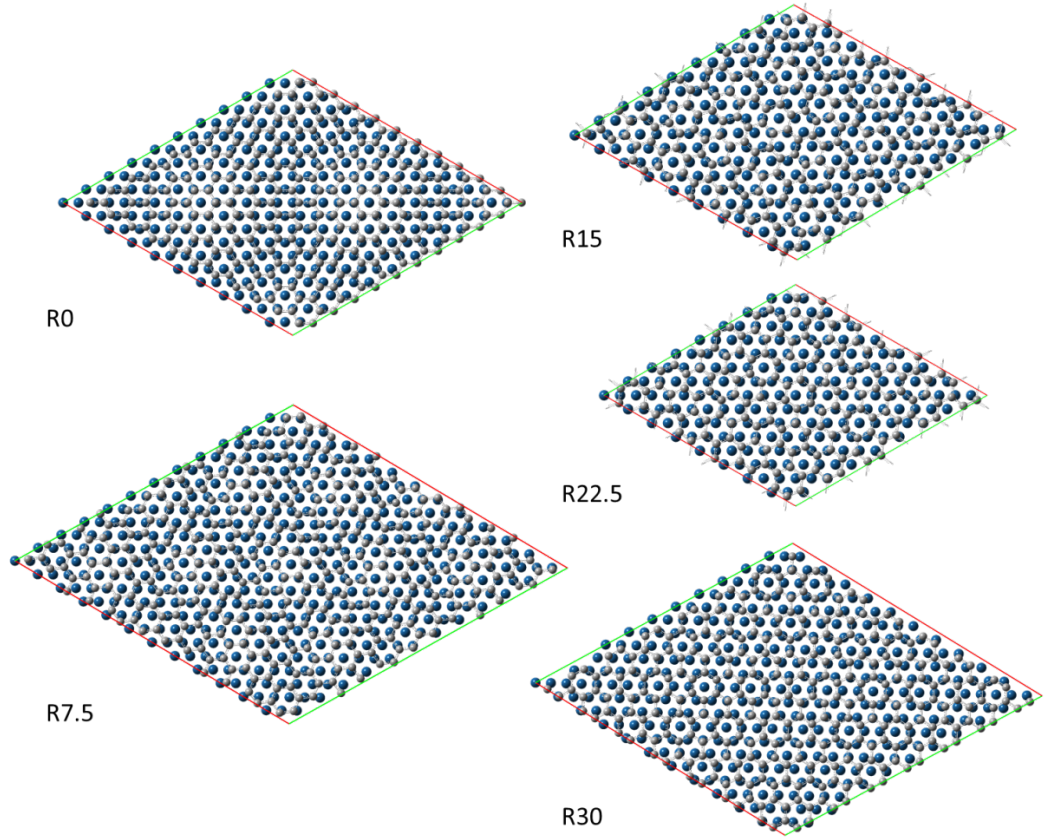

**Supplementary Figure 1: The optB86b-vdW optimized R0, R7.5, R15, R22.5, and R30 structures.**

**Supplementary Table 1: DFT Results.** The optB86b-vdW predicted graphene deformation energies (bending energies), absolute and relative binding energies and difference in the predicted values using a 1x1x1 and a 2x2x1 k-point grid ($\Delta E/\Delta k$).

|  | Graphene Deformation Energy per C Atom (meV) | | | Binding Energy per C Atom (meV) | | | Relative Binding Energy per C Atom (meV) | |
| --- | --- | --- | --- | --- | --- | --- | --- | --- |
| k-points | 1x1x1 | 2x2x1 | $\Delta E/\Delta k$ | 1x1x1 | 2x2x1 | $\Delta E/\Delta k$ | 1x1x1 | 2x2x1 |
| R0 | 0.70 | 0.63 | -0.07 | -77.11 | -76.68 | 0.43 | 0.00 | 0.00 |
| R7.5 | 0.87 | 0.88 | 0.00 | -77.82 | -77.39 | 0.43 | -0.71 | -0.71 |
| R15 | 0.23 | 0.24 | 0.01 | -76.36 | -76.48 | -0.12 | 0.75 | 0.19 |
| R22.5 | 0.10 | 0.12 | 0.01 | -77.04 | -76.62 | 0.42 | 0.07 | 0.05 |
| R30 | 0.33 | 0.33 | 0.00 | -76.93 | -76.19 | 0.74 | 0.18 | 0.48 |

**Supplementary References**